\documentclass[%
 reprint,
superscriptaddress,
 amsmath,amssymb,
 prl,
]{revtex4-1}

\usepackage{pdfpages}
\usepackage{etoolbox} 
\usepackage{pgffor}

\makeatletter
\AtBeginDocument{\let\LS@rot\@undefined}
\makeatother

\usepackage{amssymb,amsmath}
\usepackage[english]{babel}
\usepackage{tabularx}
\usepackage{soul}
\usepackage{color}
\usepackage{color, colortbl}
\definecolor{LRed}{rgb}{1,.6,.6}

\usepackage{graphicx}
\graphicspath{{./images/}}
\usepackage{array}
\usepackage{float}

\usepackage{graphicx}
\usepackage{dcolumn}
\usepackage{bm}
\usepackage{amsmath}

\usepackage{braket}

\DeclareSymbolFont{matha}{OML}{txmi}{m}{it}
\DeclareMathSymbol{\varv}{\mathord}{matha}{118}

\newcommand{\fref}[1]{Figure~\ref{fig:#1}}

\newcommand{\flabel}[1]{\label{fig:#1}}

\newcommand{\elabel}[1]{\label{eqn:#1}}

\newcommand{\vecn}[1]{#1}          
\newcommand{\X}{\vecn{x}}          
\newcommand{\XP}{\mathbf{x}}       

\graphicspath{{./}}

\begin{document}

\title{Augmented Harmonic Linear Discriminant Analysis} 

\author{Z. Faidon Brotzakis}
\affiliation{Department of Chemistry and Applied Bioscience, ETH Z{\"u}rich, c/o USI Campus, Via Giuseppe Buffi 13 Lugano, CH-6900, Lugano, Ticino, Switzerland}
\affiliation{Institute of Computational Science, Universita della Svizzera Italiana (USI), Via Giuseppe Buffi 13 Lugano, CH-6900, Lugano, Ticino, Switzerland.}
\affiliation{Contributed equally to this work}
\author{Dan Mendels}
\affiliation{Department of Chemistry and Applied Bioscience, ETH Z{\"u}rich, c/o USI Campus, Via Giuseppe Buffi 13 Lugano, CH-6900, Lugano, Ticino, Switzerland}
\affiliation{Institute of Computational Science, Universita della Svizzera Italiana (USI), Via Giuseppe Buffi 13 Lugano, CH-6900, Lugano, Ticino, Switzerland.}
\affiliation{Contributed equally to this work}
\author{Michele Parrinello}
\email{parrinello@phys.chem.ethz.ch}
\affiliation{Department of Chemistry and Applied Bioscience, ETH Z{\"u}rich, c/o USI Campus, Via Giuseppe Buffi 13 Lugano, CH-6900, Lugano, Ticino, Switzerland}
\affiliation{Institute of Computational Science, Universita della Svizzera Italiana (USI), Via Giuseppe Buffi 13 Lugano, CH-6900, Lugano, Ticino, Switzerland.}

\date{\today}



\begin{abstract}

Many processes of scientific and technological interest are characterized by time scales that render their simulation impossible if one uses  present day simulation capabilities. To overcome this challenge a variety of enhanced simulation methods has been developed. A much-used class of methods relies on the use of collective variables. The efficiency of these methods relies critically on an educated guess of the collective variables.  For this reason much effort has been devoted to the construction and improvement of collective variables. Among the many methods proposed, harmonic linear discriminant analysis has proven effective. This method builds the collective coordinates solely from the knowledge of the fluctuations in the different metastable state.   In this Letter we propose to improve upon the harmonic linear discriminant analysis by adding to the construction of the collective coordinates an extra bit of information, namely that of the transition state ensemble. Configurations belonging to the transition state ensemble  are harnessed by the use of the spring shooting transition path sampling algorithm. We show on a challenging example that these coordinates thus augmented not only perform better in the calculation of the static properties, but also accelerate considerably the calculation of reaction rates.

\end{abstract}


\maketitle



One of the main tools of contemporary science is the simulation of condensed phase systems at the atomic
level. However, in spite of the many examples of successful simulations,
several problems need to be solved in order to substantially enhance its scope allowing more and more complex systems to be studied.
One problem stands out  and it is the limited time scale of the
processes that can be simulated. This problem is made more acute by the fact
that computer technology alone will not be able to come to the rescue in the near
future.

Since very many physico-chemical phenomena take place on time scales that are
unreachable by shear computational power, several methods have been
proposed to overcome this hurdle. Here, we shall consider two
classes of methods. One class is based on the introduction of collective variables
(CV), such as umbrella sampling, metadynamics (MetaD), and variationally enhanced
sampling~\cite{Torrie1977,Laio2002,Valsson2014}. The other class instead is based on variants of the Transition Path
Sampling (TPS) approach~\cite{Dellago_1998}. These two categories are not exhaustive and many other
methods like parallel tempering or kinetic Monte Carlo exist that cannot be
classified in either of the two classes but they are not relevant in the present
context.

Methods in the first class rely heavily on an educated choice of CVs. These should
be able of approximating the slowest modes of the system. Much effort has been
devoted to designing and improving the quality of CVs\cite{Tiwary2016,Sultan2018} and a vast library of
collective variables that cover many physical phenomena is now available. Still the design of
appropriate CVs can be challenging.

Very recently much progress has been made towards the automatic
construction of CVs that can be used in the study of transitions between a given
set of metastable states. We called this method Harmonic Linear Discriminant
Analysis (HLDA)~\cite{Mendels2018}. The method is inspired by the linear discriminant analysis
(LDA) data classification approach~\cite{fisher1936use} that aims at finding the low dimensional
projection that best separates different multidimensional classes of
data. In HLDA the data to be discriminated are the configurations explored in
short molecular dynamics runs performed in the different metastable states.
A remarkable feature of HLDA is that the CVs are constructed solely from a
study of the fluctuations naturally occurring while the system is in the different
metastable states. Although the method is very recent a number of successful
applications have already been made~\cite{Mendels2018b,Capelli2019,YueYue2018,Piccini2018,Rizzi2019}.

While the realm of HLDA applications appears to be vast there are circumstances
in which this approach is expected to encounter its limits. For instance when a linear
approach is not sufficient to discriminate between basins, or when the HLDA CVs
lead to a slow convergence because some degrees of freedom relevant to the
transition are not picked up by fluctuations in the basin. In order to address this issue we will make use of some of the concepts and techniques that have been
discussed in the transition path sampling  literature. This approach focuses on the identification of
the transition paths and the notion of Transition State Ensemble (TSE). If the CVs are of
good quality then the apparent Free Energy Surface (FES) transition state indeed coincides with the TSE. As the quality of the CVs degrades this becomes less and less of an accurate statement and, although convergence can still be reached, it
is rather slow revealing that one has not fully captured the physical nature of the
transition.

For this reason we suggest that, when the HLDA procedure does not lead to fully
satisfactory CVs one can augment it by adding information on the transition state
ensemble with the help of TPS, we call this new method
Augmented Harmonic Linear Discriminant Analysis (AHLDA). This is done in steps.
First, we perform a standard HLDA calculation. From the reactive trajectories we
extract configurations that are used as a seed for  path sampling in a
spirit similar to that of refs~\cite{Borrero2016,Mendels2018b}. Using the spring shooting transition path sampling algorithm~\cite{Brotzakis2016a} we identify a sufficiently large number of configurations belonging to the transition state
ensemble. This set of configurations is added to the classes to be analyzed by
HLDA and a new set of CVs, that encodes information not only
on the metastable states but also on the transition states, is generated. This has several
advantages, it leads to a more efficient exploration of the free energy space, it
identifies unambiguously the transition state ensemble, and allows, as we shall
see, a more efficient rate calculation.

One could also reverse this point of view and look at AHLDA not as a method to
improve the search for CVs to be used in MetaD and related methods, but
as a way of improving TPS based methods by offering a way of computing free energies in a much
easier way than what has so far been proposed in the literature~\cite{Rogal2010a}.

\section*{Methods}

We first review the multi-class HLDA method. One first needs to define a set of descriptors $d_i(\mathbf{R})$ $i=1,...,N_d$ capable of identifying the different metastable states. For each metastable state $I$ we compute the $N_d$ dimensional vector of the average values of the descriptors $\mu_I$ and the fluctuation matrix $\Sigma_I$. Given a set of $M$ classes the distribution of the resulting projection is $M - 1$ dimensional. In the $d_i$ space we look for the matrix $\mathbf{W}$ that when applied to the vector $d_i$ produces $M-1$ directions such that when projected onto these variables the original multidimensional distributions are best separated. The matrix $\mathbf{W}$ is computed by maximizing a Rayleigh-like ratio
 
\begin{equation}
\label{fisher_ration}
\mathcal{J(\mathbf{W})} = \frac{\mathbf{W}^T \mathbf{S}_b \mathbf{W}}{\mathbf{W}^T \mathbf{S}_w \mathbf{W}}
\end{equation}

\noindent where the between matrix $S_b$ is a measure of the distance between the projected classes given by 

\begin{equation}
\label{Eq:HLDASb}
\boldsymbol{S_b}=\sum_{I}^{M}\left( \boldsymbol{\mu}_I -\boldsymbol{\overline{\mu}})(\boldsymbol{\overline{\mu}}-\boldsymbol{\mu}_I \right)^T
\end{equation}

\noindent where $\overline{\mu}$ is the overall mean of the data sets, i.e.  $\overline{\mu}=\frac{1}{M}\sum_{I}^{M}\mu_I$. The measure of the overall spread of the projection is instead given by $S_w$ that in HLDA is considered by the harmonic average of the fluctuation covariance matrices

 \begin{equation}
\label{Eq:HLDASw}
\boldsymbol{S_w}=\frac{1}{\frac{1}{\boldsymbol{\Sigma}_A}+\frac{1}{\boldsymbol{\Sigma}_B}+ ...+ \frac{1}{\boldsymbol{\Sigma}_M}}
\end{equation}

\noindent This  amounts at taking as a measure of the total spread the harmonic average of the spreads. We note that in standard LDA $\boldsymbol{S_w}=\boldsymbol{\Sigma}_A+\boldsymbol{\Sigma}_B+ ...+ \boldsymbol{\Sigma}_M$. This choice of using the harmonic average has some chemical as well as Bayesian justification. The maximization of $\mathcal{J(\mathbf{W})}$ is obtained on the solution of the eigenvalue equation: 

\begin{equation}
\label{Eq:HLDAeig}
\boldsymbol{S_w^{-1} S_b W}=\lambda \boldsymbol{W} 
\end{equation}

\noindent The $M - 1$ lowest eigenvectors define the optimal directions and are used as CVs.

\noindent

In order to generate an ensemble of configurations centred around the transition state it seemed
natural to us to use the spring shooting algorithm of Brotzakis and Bolhuis~\cite{Brotzakis2016a}. In this approach a
sequence of one sided shooting points along the trajectories are generated through a Monte
Carlo procedure that ensures that the shooting points density peaks around the transition state.

Given a trajectory $\XP^{(n)}$ generated after a one sided shooting at time $\tau$ a new shooting point $\tau'$ is
generated with probability:

\begin{equation}
\label{Eq:TPS_prob}
p(\tau)=Ce^{sk\tau}
\end{equation}

\noindent where $s$ is chosen with equal probability to be 1 or -1. The sign of $s$ determines whether the new shooting move should be in the forward (1) or backward (-1) direction, and the constant $k$ determines the width of the distribution $p(\tau)$. In the practice in order to sample this distribution one sets up a Monte-Carlo procedure based on the acceptance ratio

\begin{align}
P(\tau_n \rightarrow \tau_{n+1}) = \min \left[1, \frac{ \exp ( s k \tau_{n+1} )  }{\exp (s k \tau_n) } \right]
\label{selectcrit}
\end{align}

\noindent One then shoots forward or backwards according to the sign of $s$ and accepts the trajectory if  it successfully reaches the forward or backward basin. The new segment of the trajectory is glued to the old one leading to a new trajectory $\XP^{(n+1)}$. This whole procedure can be summarized in the acceptance probability

\begin{flalign}
	\begin{split}
P_{acc}[\tau_n \rightarrow \tau_{n+1} ; \XP^{(n)}\rightarrow
                 \XP^{(n+1)}] = \\ h_A(\X_0^{(n+1)}) 
              h_B(\X_L^{(n+1)})  \min \left[1, \frac{ \exp ( s k \tau_{n+1} )  }{\exp (s k \tau_n) } \right]
\elabel{selectcrit_full}
	\end{split}
\end{flalign}

\noindent where the indicator functions $h_A(\X_0^{(n+1)})h_B(\X_L^{(n+1)})$ ensure that the initial $\X_0^{(n+1)}$ and final $\X_L^{(n+1)}$ point of the of the trajectory $\XP^{(n+1)}$  lie in basins A and B respectively. This algorithm has been used with success to uncover TSEs of complex bio-molecular transitions~\cite{Brotzakis2019}.

To enhance the sampling of the system of interest using our CVs we utilize MetaD ~\cite{Laio2002}. MetaD accelerates  sampling by adding a history-dependent bias in the form of Gaussian kernels on the selected CVs. In well tempered MetaD~\cite{Bonomi2010} this aim is achieved by periodically adding a bias that is updated according to the iterative procedure

\begin{equation}
\elabel{eq:bias}
V_n(\bold{s})=V_{n-1}(\bold{s})+G(\bold{s},\bold{s}_{n}) \exp \left( -\frac{1}{\gamma-1} V_{n-1}(\bold{s}_n) \right)
\end{equation}

\noindent where V$_n$($\bold{s}$) is the total bias deposited at iteration $n$ and is obtained by adding at the previous bias V$_{n-1}(\bold{s}$) a contribution that results from the product between a Gaussian kernel $G(\bold{s},\bold{s}_{n})$ and a multiplicative factor $\exp \left( -\frac{1}{\gamma-1} V_{n-1}(\bold{s}_n)\right)$ that makes the height of the added Gaussian diminish with time. The bias factor $\gamma >$ 1 determines the rate with which the added bias decreases and regulates the amplitude of the $\bold{s}$ fluctuations. At convergence 

\begin{equation}
\label{Eq:FES_metaD}
F(\bold{s})=-(1-\frac{1}{\gamma})V(\bold{s})
\end{equation}

\noindent From the MetaD trajectory the expectation value of any operator $O(\bold{R})$ can be  calculated using the reweighting procedure of Tiwary and Parrinello~\cite{Tiwary2015c}:

\begin{equation}
\label{Eq:Reweighting}
F(\bold{s})=-\left< O(\bold{R}) e^{\beta(V(s(\bold{R},t))-c(t))} \right>
\end{equation}

\begin{figure}[th!]
\includegraphics[width=6.6cm]{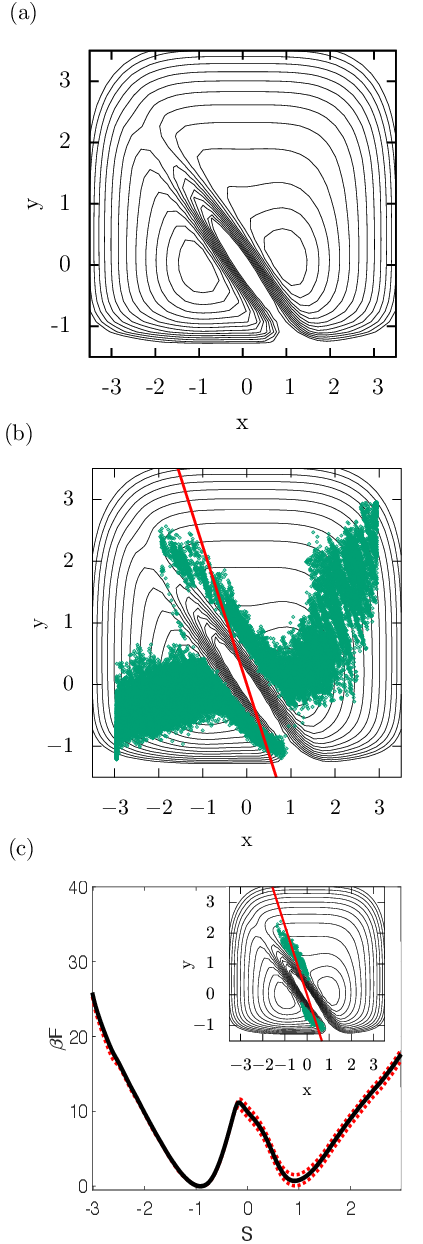}
\caption{\small \flabel{Fig1} a) Potential, with contour frequency of 0.3 $\beta V$.  b) scatter plot of a MetaD simulation using the HLDA CV, $s=0.91x+0.41y$ where in in red is the separating hyper-plane obtained using HLDA. c) FES projection on $S$ and the respective uncertainty obtained using the reweighting algorithm of ref.~\cite{Tiwary2015c}. In the inset we project on x, y plane the configurations corresponding to the barrier region of the $S$ projection.}
\end{figure}

\noindent where the time dependent energy offset $c(t)$ is given by 

\begin{equation}
\label{Eq:c of t}
c(t)=-\frac{1}{\beta}\frac{\int ds e^{-\beta(F(\bold{s})+V(\bold{s},t))}}{\int ds e^{-\beta F(\bold{s})}}
\end{equation}

\noindent Since our procedure has identified the TS region and our CVs are able to discriminate between the TS and the metastable states, we want to use this property to improve methods like infrequent MetaD and the variational flooding~\cite{Tiwary2013,McCarty2015}. We recall that these  methods are derived from the Hyperdynamics of Voter~\cite{Voter1997}  or the potential flooding of Grubmuller~\cite{Grubmuller1995}. The basic idea in this class of approaches is that in a rare event scenario the escape time $\tau_M$ from a metastable state that occurs in a biased simulation  is translated to the physical one $\tau$ by the relation 

\begin{equation}
\elabel{eq:acc_fact}
t=t_M e^{\beta V(\mathbf{s},t_{M})},
\end{equation}

\noindent provided that at $\tau_M$ no bias has been added to the TS region. The calculation is repeated several times for each metastable state. In a rare event scenario the distribution of $\tau$ should be Poissonian, a statement whose accuracy can be verified using a Kolmogorov-Smirnov  test~\cite{Salvalaglio2014}. In infrequent MetaD one satisfies the condition that the TS region is not contaminated by the bias by reducing the frequency with which the Gaussians are deposited. In variationally enhanced sampling a different strategy is applied to achieve this result but for the sake of space we will not discuss it here, even if also here the results of our approach could be applied with profit.

\subsection*{Example}
 
We illustrate the benefits of AHLDA by simulating a particle that moves with Langevin dynamics on the potential energy surface shown in  \fref{Fig1}a. The interested reader can find details about the potential and the corresponding dynamics and MetaD simulations in the supplemental material. This potential exhibits two minima separated by a high ridge with two transition states at (-1.5,1.9) and (1,-1.1)  lying at the foot of the ridge. The lowest TS is the first one. The second is at high energy and at least at low temperatures does not affect the rate of transition from one well to another. The minima A and B are at positions (-1,0) and (1.1,0) respectively. A key feature of this potential that makes it challenging is that the principal components in the minima lie approximately parallel to the ridge of the energy barrier. In fact, this potential is often used as a testing ground for enhanced sampling methods and mimics a number of physical/chemical processes~\cite{Bolhuis2002}.

We first performed HLDA calculations and found the CV, whose perpendicular hyperplane is shown in \fref{Fig1}b. It is seen that the hyperplane is almost parallel to the ridge, yet it still mixes the barrier with the metastable states. Thus it is not very efficient in accelerating transitions from one well to the other. It is still able to induce well to well transitions but the rate of convergence is painfully slow. In particular, calculating $\beta F(\bold{s})$ directly from the bias as seen in Figure S1 shows that even after $10^9$ time steps a satisfactory result has not yet been obtained. A faster convergence can be obtained if we estimate $\beta F(\bold{s})$ not directly from the bias but rather using the reweighting procedure of Eq.~\ref{Eq:Reweighting} as shown in ~\fref{Fig1}c. This discrepancy is a sign of a bad CV. Another sign of a poor CV is that the apparent $\beta F(\bold{s})$ TS does not overlap with the dynamical TS. Moreover, the barrier height is different from the value $\beta \Delta F = 1.9$ which is the potential energy surface barrier height.
This calculation has two sides to it. On the one hand it shows that even if it struggles, HLDA does give reasonable if not perfect results. On the other, it shows that important information is being missed. Also, given the fact that the apparent TS is contaminated, it would be foolhardy to attempt infrequent MetaD or variationally enhanced sampling flooding to compute rates. 

\begin{figure}[t!]
\centering
\includegraphics[width=6.7cm]{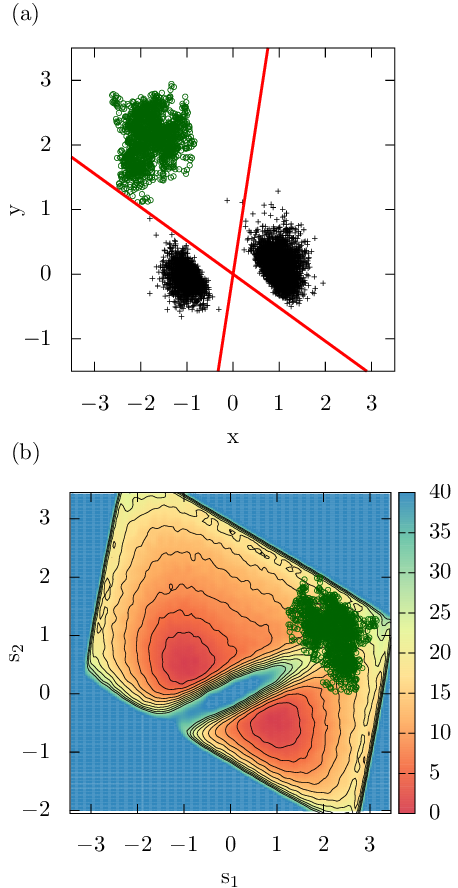}
\caption{\small \flabel{Fig2}a) AHLDA separatrices corresponding to CVs s$_1$ and s$_2$ and metastable state fluctuations (black) and transition state ensemble (dark green). b)  $\beta F$ at T=0.1 as a function of  $s_1=-0.98x+0.21y$ and  $s_2=0.46x+0.89y$. In dark green the TSE points.}
\end{figure}

Given the poor performance of HLDA, we now apply our recipe to improve upon it. To this effect we perform a spring shooting TPS calculation starting from one of the trajectories generated during the HLDA based MetaD simulations. The details of the spring shooting calculations are described in the supplemental material. At the end of this procedure a TSE centered class of states was found (see \fref{Fig2}).
By adding the TSE class to the existing ones corresponding to the metastable states, we thereby augment HLDA.  AHLDA provides with two CVs, whose corresponding perpendicular planes are plotted in \fref{Fig2}a. It can be seen that not only the two metastable states are separated but also the TS set of states can be well discriminated. A MetaD simulations is performed using these new CVs, and now a steady convergence is achieved and the $V(\bold s)$ based FES estimate (Eq. \ref{Eq:FES_metaD}) and the reweighting (Eq.\ref{Eq:Reweighting}) one are now in agreement with one another. Also, the apparent TS in the augmented HLDA 2D surface does coincide with the dynamical one (see~\fref{Fig2}b).


\begin{figure}[t!]
\centering
\includegraphics[width=6.7cm]{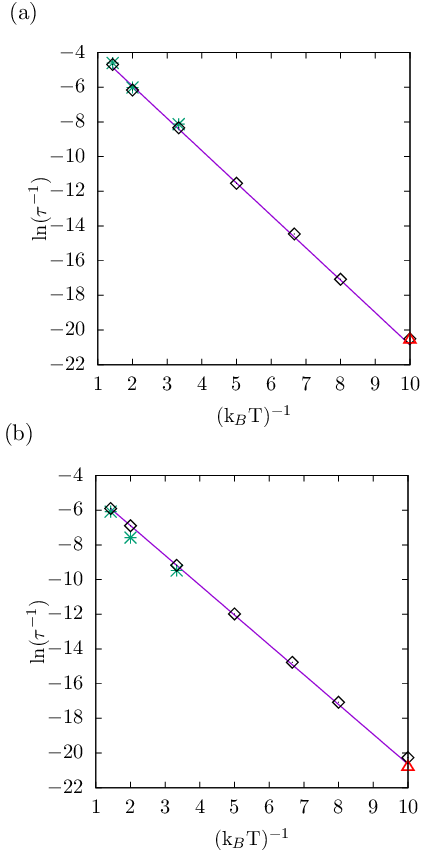}
\caption{\small \flabel{Fig3}  Arrhenius plot for the transition rates calculated for a) the $A \rightarrow B$  and b) the $B \rightarrow A$ transitions using infrequent MetaD (black diamonds), unbiased simulations (green crosses) and MetaD (red triangles).}
\end{figure}

Finally, we assess the ability of the AHLDA CVs to accurately capture the system transition rates. We do this by using infrequent MetaD. For educational purpose before doing this we perform infrequent MetaD calculations at T=0.1 $k_BT$ using the HLDA CV. As expected the calculation of dynamical properties proved even more challenging than the static ones. In these calculations, a large fraction of transitions did not occur through the system TS  and the escape times were only marginally Poissonian distributed. In contrast, the AHLDA CVs $s_1$ and $s_2$ proved ideal for this purpose, and the statistics of the escape times were found to give rise to high p-values in the Kolmogorov-Smirnov test. The rates (A $\rightarrow$ B and B$\rightarrow$ A) obtained by infrequent MetaD with these CVs (see~\fref{Fig3}) show an Arrhenius behaviour, and at high temperatures they agree with unbiased estimates. In addition, the activation energy obtained from the Arrhenius plot slopes is  ~1.82 $\pm$ 0.2, in agreement with the analytical estimate of the energy at the top of the barrier of $\beta$V = 1.9. 

A central aspect of the use of infrequent MetaD is that the deposition of bias is done sufficiently slowly such that none is added to the system TS. This evidently is done since generally the system TS location in not known. In the present case in contrast, given that this position is known a priori we tested the possibility of increasing the deposition rate in the rate calculation simulations while preventing bias to be deposited on the TS region by default. Thus, we gradually increased the Gaussian deposition frequency in the rate calculation simulations and found that these can be increased by more than a factor of three with respect to the standard deposition rate used in infrequent MetaD as exemplified by the result obtained for T=0.1 $k_BT$ (see~\fref{Fig3}). 



\subsection*{Conclusions}

Constructing effective CVs for complex processes can be a highly challenging task, often constituting the de facto solution of the problem being considered. In this Letter we presented AHLDA for the automated construction of CVs. AHLDA rests on using information contained in the TSE along with that contained in the fundamental fluctuations of metastable states of the system. Using such CVs in MetaD simulations of a prototypical test system characterized by an important degree of freedom which is not apparent in the metastable states and by nonlinear transition pathways, enabled fast convergence and lead to a physically and dynamically meaningful description of the system. 
An area in which our approach appears to be particularly promising is in the calculation of rates. Further efficiency improvements can be envisioned by using more advanced rate calculation techniques.

{\color{white}sfs\par}

The authors thank Prof. Peter Bolhuis for providing the analytical form of this potential and Dr. David Swenson, Arjun Wadhawan,  Luigi Bonati and Michele Invernizzi for helping in the the setup of the implementation of  the simulations.  This research was supported by the VARMET European Union Grant ERC$-$2014$-$ADG$-$670227. Computational resources were provided by the Swiss National Supercomputing Centre (CSCS).

\bibliography{library}



\section*{Supplementary material}

# Supplemental Materials: Augmented Harmonic Linear Discriminant Analysis

Z. Faidon Brotzakis,[1,2,3] Dan Mendels,[1,2,3] and Michele Parrinello[1,2,*]

[1]*Department of Chemistry and Applied Bioscience, ETH Zürich, c/o USI Campus, Via Giuseppe Buffi 13 Lugano, CH-6900, Lugano, Ticino, Switzerland*

[2]*Institute of Computational Science, Universita della Svizzera Italiana (USI), Via Giuseppe Buffi 13 Lugano, CH-6900, Lugano, Ticino, Switzerland.*

[3]*Contributed equally to this work*

(Dated: February 23, 2019)

## COMPUTATIONAL DETAILS

### Toy model potential

The simulations were performed in Langevin dynamics with mass $m = 1$, timestep $\Delta t = 0.005$ time units and friction $\gamma = 10$ (time units)$^{-1}$. This potential encounters two stable states separated by a high energy ridge, with two pertinent transition states at positions (-1.5,1.9) and (1,-1.1) respectively. The first is the lowest TS. The second is at high energy and at least at low temperatures does not affect the rate of transition from one well to another. The minima A and B are at positions (-1,0) and (1.1,0) respectively. We used the Wolfe-Quapp potential implemented in PLUMED [1] to construct the expression of the potential in Eq. 1.

$$\beta V(x,y) = -e^{-((x-1)^2+y^2)} - e^{-((x+1)^2+y^2)} + 5e^{-0.32(x^2+y^2+20(x+y)^2)} + \frac{32}{1875}(x^4+y^4) + \frac{2}{15}e^{-2-4y} \quad (1)$$

### Metadynamics

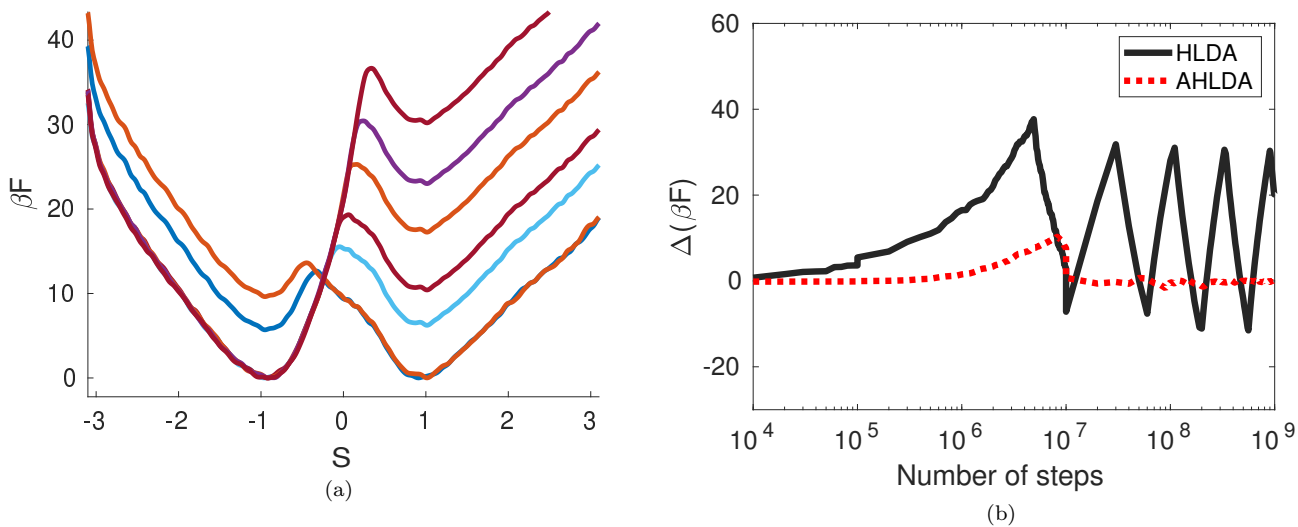

Figure S1: a) Free energy profile as function of s, calculated directly from the simulation bias (eq. 9) for consecutive simulation times with an interval of $t = 6 \times 10^6\, steps$. b) Free energy difference between the two metastable states as a function of time for the HLDA (black) and AHLDA (red) simulations calculated directly from the bias as in equation 9 of the main text.

We used the Well Tempered variant of Metadynamics [2] in natural units, with hill height 1 $k_BT$, where $k_B = 1$, biasfactor 40, sigma 0.05 and steps/deposition of 10000 $\Delta t$. For the rate calculations we performed standard infrequent Metadynamics and Metadynamics where care was taken not to add Gaussians whenever the system was in the transition state region as estimated by the spring shooting algorithm. In the first simulations we used a hill height of 1 $k_BT$, bias steps/deposition of 10000 $\Delta t$ and a biasfactor spanning from 7 to 40, whereas in the second, we used a hill height of 1 $k_BT$, biasfactor of 7 and bias steps/deposition of 3000 $\Delta t$. In both simulations we used a sigma of 0.05.

### Transition Path Sampling

States are defined as circles centered at positions (-1.5, 0) and (1.5, 0) and of radius $r = 0.45$. Frames were saved at frequency of 10 $\Delta t$. The spring shooting constant $k$ is set to the value of $k = 0.1$ and $\Delta\tau_{max} = 20$ frames. We performed a total of 2000 MC cycles with acceptance ratio of 28% and 200 decorrelated transition pathways. The TSE was obtained from the Least Changed Path analysis (LCP) as done in Refs. [3, 4]. This part of the reactive

trajectories is the part of the path ensemble that is the least changed during the path sampling and serves as an approximation of the TSE. The path tree in Fig. S2a, shows that despite the asymmetry of the barrier, the algorithm can sample both forward and backward paths offering a good decorrelation.

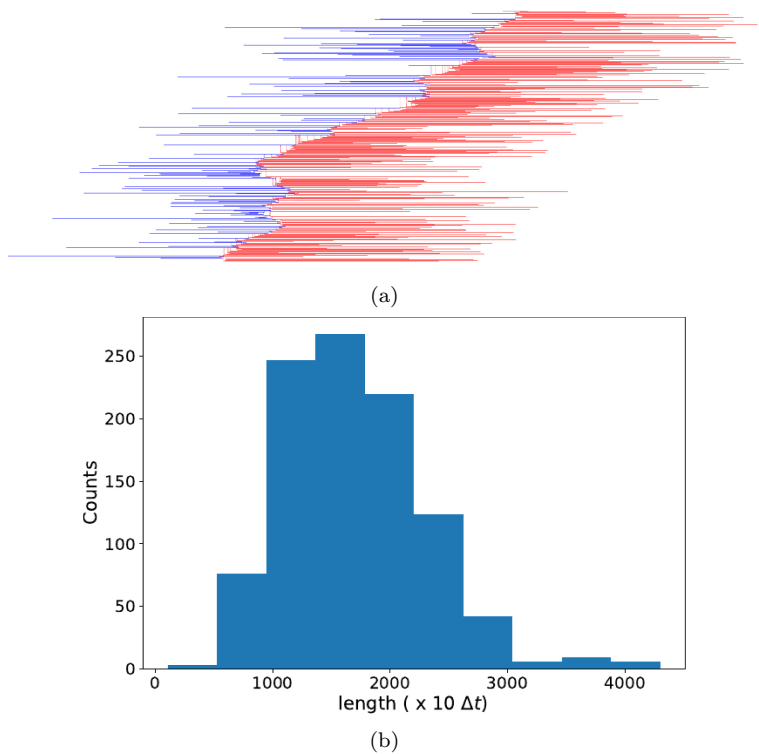

(a)

(b)

Figure S2: a) Path tree. With blue and red are indicated the lengths of the backward and forward partial paths respectively. The vertical lines indicate the position of the shooting points. b) Path length histogram of the transition path ensemble.

### Rate estimate as a function of bias deposition frequency

In Figure S3 we show the escape time and p-value estimate as a function of the steps/deposition . The escape time is overestimated high deposition pace and asymptotically-after deposition time of 3000- converges. A similar picture holds for the p-value, which is severely below the threshold of 0.05 at high deposition pace and gradually, as the deposition time increases-after deposition time of 3000-, the statistics can be trustworthy.

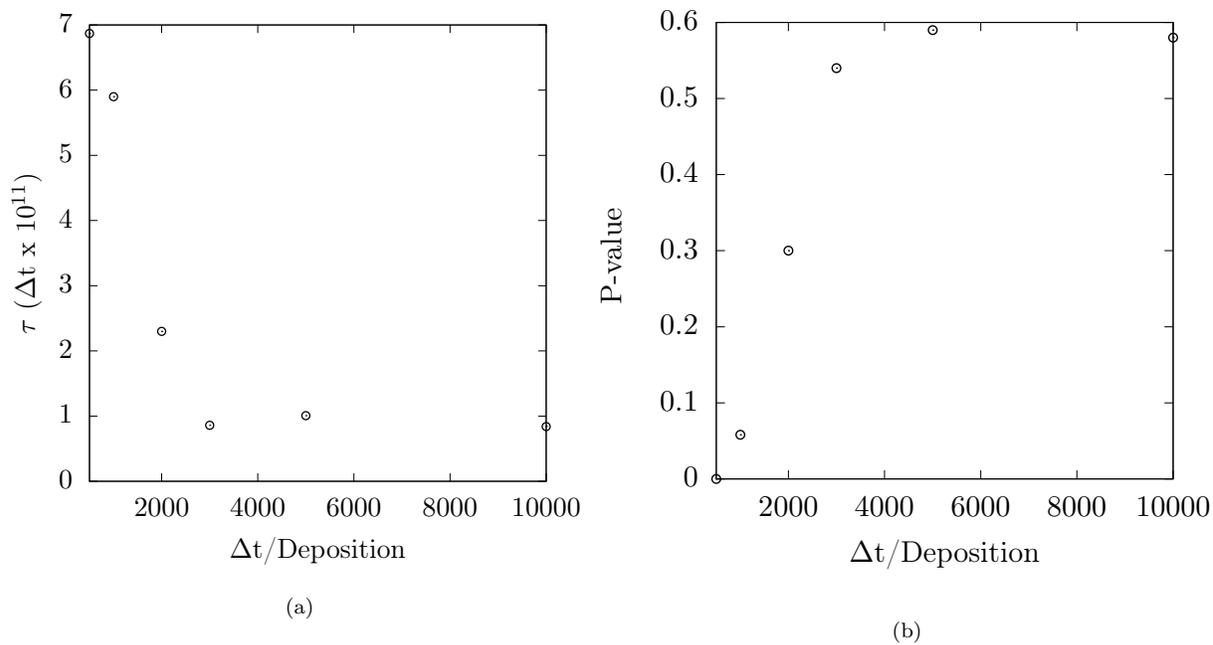

Figure S3: a)The escape time and b) p-value, estimated as a function of the deposition time



\end{document}